
\input phyzzx
 \hsize=15.8cm
\vsize=23cm

\input epsf
\ifx\epsfbox\UnDeFiNeD\message{(NO epsf.tex, FIGURES WILL BE IGNORED)}
\def\figin#1{\vskip2in}
\else\message{(FIGURES WILL BE INCLUDED)}\def\figin#1{#1}\fi

\def\ifig#1#2#3{\xdef#1{fig.~\the\figno}
\goodbreak\midinsert\figin{\centerline{#3}}%
\smallskip\centerline{\vbox{\baselineskip12pt
\advance\hsize by -1truein\noindent\footnotefont{\bf Fig.~\the\figno:} #2}}
\bigskip\endinsert\global\advance\figno by1}

\def\footnotefont{\tenpoint}

\newwrite\ffile\global\newcount\figno \global\figno=1
\def\fig{fig.~\the\figno\nfig}
\def\nfig#1{\xdef#1{fig.~\the\figno}%
\writedef{#1\leftbracket fig.\noexpand~\the\figno}%
\ifnum\figno=1\immediate\openout\ffile=figs.tmp\fi\chardef\wfile=\ffile%
\immediate\write\ffile{\noexpand\medskip\noexpand\item{Fig.\ \the\figno. }
\reflabeL{#1\hskip.55in}\pctsign}\global\advance\figno by1\findarg}

\parindent 25pt
\overfullrule=0pt
\tolerance=10000
\def\Re{\rm Re}

\def\half{{\textstyle {1 \over 2}}}

\def\ie{{\it i.e.}}

\nopagenumbers
\baselineskip=14pt

\hfill {DAMTP/94-25}\break
 \null
\vskip 5cm
\centerline{VACUUM VALUES FOR AUXILIARY STRING FIELDS.}
\vskip 1cm
 \centerline{ Michael B.  Green,}
\centerline{DAMTP, Silver Street, Cambridge CB3 9EW, UK\foot{email:
M.B.Green@amtp.cam.ac.uk}}
\nopagenumbers
\vskip 4cm
\abstract
Auxiliary string fields are introduced in light-cone gauge string field theory
in order to express
contact interactions as contractions of cubic vertices.    The
auxiliary field in the purely closed-string  bosonic theory may be given a
non-zero expectation value, giving rise to a  phase in which
world-sheets have boundaries.
\vfill\eject

\pagenumbers
\sequentialequations

The sum over Riemann surfaces that  represents the Feynman diagrams of
perturbative string theory can be obtained from  light-cone gauge string field
theory  \REF\mandelstama{S.  Mandelstam, {\it Interacting-string picture of
dual resonance models}, Nucl.  Phys.  {\bf B64} (1973) 205.}\REF\kakua{K.
Kikkawa and M.  Kaku, {\it Field theory of relativistic strings.  I.  Trees},
Phys.  Rev.  {\bf D10} (1974) 1110; {\it Field theory of relativistic strings.
II.  Loops and pomerons}, Phys.  Rev.  {\bf D10} (1974)
1823.}\REF\gervaisa{J-L. Gervais and E.  Cremmer, {\it Combining and splitting
relativistic strings},  Nucl.  Phys.  {\bf B76}   (1974) 209.}
[\mandelstama-\gervaisa].    In the first part of
this paper the stringy Feynman rules will be reexpressed by the inclusion of
auxiliary string fields which clarify  the occurence of  contact
interactions  in  both open and closed string theory.  Furthermore,  it will be
argued that in the
 bosonic closed-string theory the auxiliary field may be given an expectation
value consistent with Poincar\'e symmetry, thereby significantly altering the
string vacuum
state.  The resulting theory is one that is described by a sum over
world-sheets with boundaries on which the embedding coordinates are constant
and with no net momentum flowing through any given boundary \REF\greena{M.B.
Green, {\it Point-like  structure and off-shell dual strings},  Nucl. Phys.
{\bf B124} (1977) 461;   {\it Modifying the bosonic string vacuum},
Phys. Lett. {\bf 201B}  (1988) 42.} \REF\greenaa{M.B.  Green,  {\it Space-time
duality and Dirichlet string theory}, Phys.  Lett. {\bf 266B} (1991)
325.}   ([\greena,\greenaa] and references therein describe some properties of
the resulting theory).  Such boundary insertions can also be formulated
covariantly but it is only in the light-cone parametrization that they are
described by an instantaneous interaction term in the hamiltonian.  A related
scheme might apply to open-string theory but
whether similar considerations can be applied  to  superstring theories is not
obvious.

In light-cone gauge string field theory the functional string field
depends on the transverse components of the string coordinates, $X^i(\sigma),
(i=1, \dots, D-2$), as well as $p^+$ and $\tau = X^+$ (the light-cone \lq
time').  The parameter $\sigma$ has  range $\pi \alpha = 2\pi p^+$. The string
field may be expressed (at $X^+=0$) in terms of the Fock space states by
$$\Phi[X] \equiv \Phi[X^i(\sigma), \alpha]= \langle X
|\Phi\rangle.\eqn\fockspaa$$
The free string hamiltonian may be written in this notation as
$$\eqalign{H_2  & = \int DX^i d\alpha \Phi [X^i, -\alpha] \int_0^{\pi\alpha}
d\sigma \left( - {\delta^2 \over \delta X^{i2}} + {1\over \pi^2} X^{\prime 2}
\right) \Phi [X, \alpha ] \cr 
& = \langle V_2|\Phi_1\rangle |\Phi_2\rangle \equiv \langle V_2|X_1\rangle
|X_2\rangle \ \Phi[X_1] \Phi[X_2], \cr}\eqn\bloker$$
which defines the string propagator.  In the light-cone parametrization Riemann
surfaces are constructed from flat sectors of world-sheet by joining
vertices with propagators, all the curvature being located at discrete \lq
interaction points'.  A similar parametrization arises in the \lq
light-cone-like'  version of covariant string field theory \REF\siegela{W.
Siegel, {\it Covariantly second-quantized string.  II. }, Phys.  Lett.  {\bf
151B} (1984) 391:  {\it Covariantly second-quantized string.  III. }, Phys.
Lett.  {\bf 151B} (1984) 396.}\REF\hikkoa{H. Hata, K. Itoh, T.  Kugo, H.
Kunitomo and K.  Ogawa, {\it  Covariant string field theory}, Phys.  Rev. {\bf
D34} (1986) 2360.}\REF\neveua{A.  Neveu and P.C.  West, {\it The interacting
gauge covariant bosonic string}, Phys.  Lett. {\bf 168B} (1986) 192.}
[\siegela--\neveua], in which $\alpha$ is an arbitrary parameter.  Many of the
issues discussed below have covariant generalizations within that framework.

The usual cubic interaction vertex is defined as the overlap between two
incoming strings  and the outgoing string (or the time-reversed process),   
$$\eqalign{H_3  & = \int \left(\prod_{r=1}^3 DX^i_r d\alpha_r \right) \Phi
[X_1,
\alpha_1]  \Phi [X_2, \alpha_2]  \Phi [X_3, \alpha_3]  
\delta [X_1,X_2,X_3] \delta (\sum_{r=1}^3 \alpha_r)  \cr &
= \langle V_3  |\Phi_1\rangle |\Phi_2\rangle |\Phi_3\rangle  \equiv   \langle
V_3  |X_1\rangle |X_2\rangle|X_3\rangle \  \langle X_1|\Phi_1\rangle \langle
X_2 |\Phi_2\rangle\langle X_3 |\Phi_3\rangle  
 , \cr}\eqn\intnorm$$
where $\delta[X_1,X_2,X_3]$ is the functional delta function that imposes the
condition that the transverse coordinates are continuous at the vertex
(and the notation implies integration over the coordinates of the complete sets
of intermediate states).   This defines the vertex, $\langle V_3|$,   as a
state in the tensor product of the Fock spaces of the three interacting
strings.  This structure applies equally to open and closed strings and also to
string theories with a richer world-sheet structure  such as superstring
theories (generally the vertex also includes  an operator insertion at the
interaction point that does not concern us here).

One way of obtaining the vertex is to consider the functional integral for the
process in which three  strings are in arbitrary asymptotic Fock space states
and propagate from $\tau=\pm \infty$.  The external legs are then lopped
off to give the vertex, in the usual manner.   However, this is not the whole
story since some regions of the moduli space of multi-particle tree amplitudes
are not reproduced by sewing such cubic vertices together.  Although these
regions are of zero measure   in the bosonic closed-string case there are
circumstances in which their presence can be important.  

\noindent{\it Auxiliary string fields and contact terms}\hfill\break\indent
To see that there are other contributions to the string interaction, consider
the process in which legs 1 and 2 of the vertex couple to definite  asymptotic
on-shell
Fock states, $|n_1\rangle$ and $|n_2\rangle$ (where the labels summarize the
occupation numbers of all the oscillator states), with widths $ \pi \alpha_1$
and  $\pi \alpha_2$, while the third (of width $\pi \alpha_3 = -\pi
\alpha_1-\pi \alpha_2$) is attached to a propagator that propagates a finite
time to the state $|X_3\rangle$ at $\tau=T$.  
The amplitude defined by this process is given by (to be specific  the
following argument will refer to  a purely closed-string theory)
$$I (1,2,X_3;T ) =\langle V_3|  n_1 \rangle |n_2\rangle |X_3' \rangle
\Delta(X_3',X_3;T),\eqn\procesr$$
with
$$\Delta(X_3',X_3;T ) =  {1\over \alpha_3}  \langle X_3'|e^{P^- T}|X_3\rangle,
\eqn\propdef$$
where $P^-$ is the hamiltonian in the light-cone frame and is given in terms of
modes (for closed strings) by $\alpha P^- = p^{i2} + 2N + 2\tilde N -4$, where
$N$ and $\tilde N$ are the level numbers of the right-moving and left-moving
modes, respectively.

\ifig\fone{(a)  The string world-sheet for the case $\eta > 0$.  The boundary
(represented by the dashed line at  $\Re \rho=\tau =0$)  of width
$\pi(\alpha_1+\alpha_2)$ represents the string state  $|X_3\rangle$ and the
interaction time is the modulus, $T$.     Dotted lines are to be identified to
form cylinders, as indicated by the labels.  (b)  When $\eta<0$ there are two
kinds of diagrams.  In (bi)  (where $q>-\eta$)  the boundary  occurs at the end
of a cylindrical portion of world-sheet (the length of the cylinder is a
modulus).  The second contribution is shown in (bii) (where $q<-\eta$)  in
which the boundary has two turning points.  The distance between the turning
points is  the  modulus, $R$.   (c)  When $\eta=-1$ ($\alpha_1 = -\alpha_2$)
only the second kind of diagram in (b) contributes.  The boundary now
represents a slit in the world-sheet at a fixed value of $\Re \rho$.}
{\epsfbox{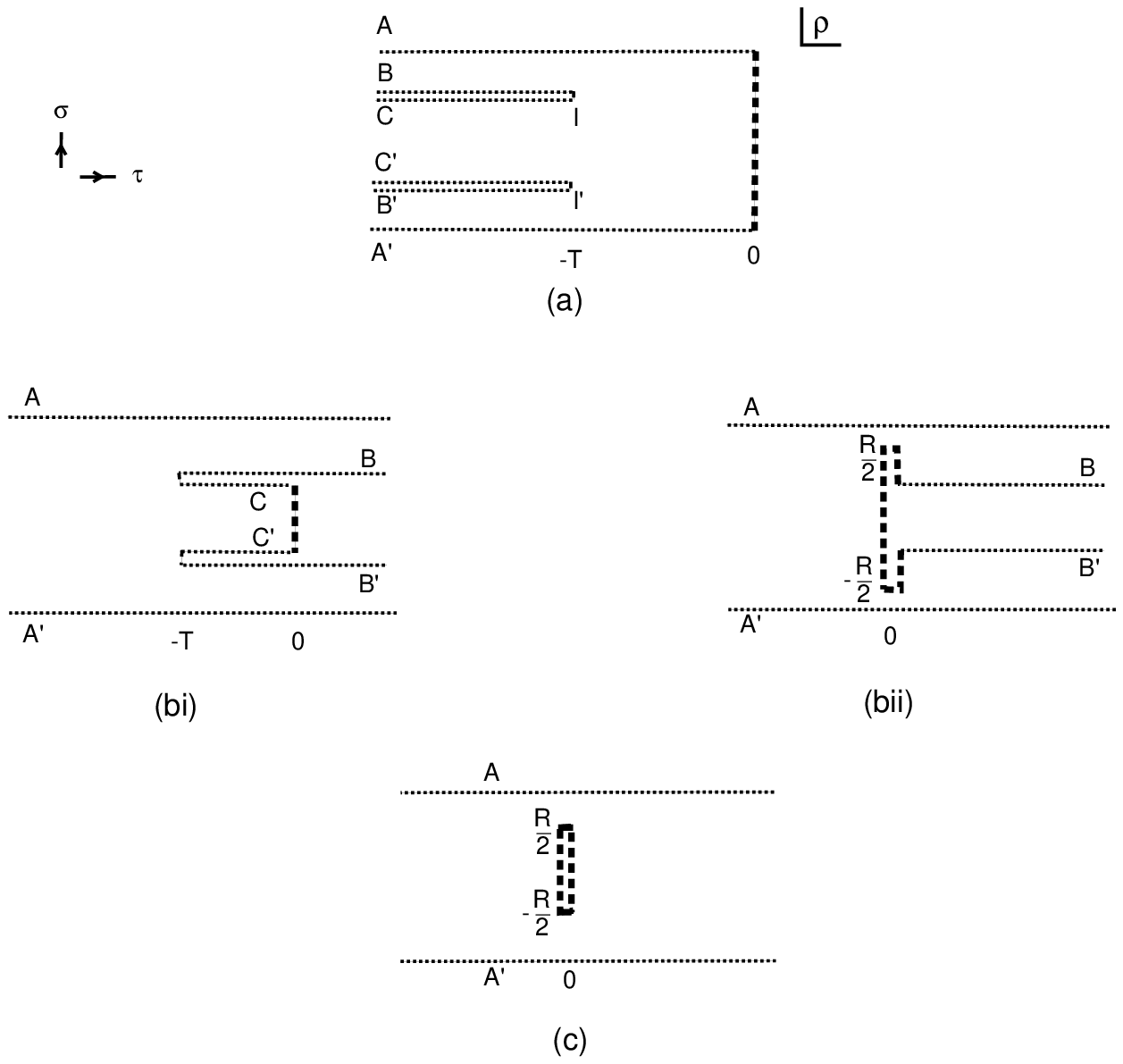}}
The process described by  $I$ defines a world-sheet with a boundary which can
conveniently be mapped to the upper-half $z(=x+iy)$-plane,  with the $x$ axis
being the image of the string
$X_3(\sigma)$.    Strings 1 and 2 are mapped to two infinitesimal holes (\ie,
punctures) that may
be located at $z=i$ and $z=iq$ (where $q$ is the real modulus and $0\le q \le
1$) with no loss
of generality.   The mapping from the string
diagram to the upper-half plane is given by
$$\rho = {\alpha_1\over 2} \ln\left(z-i\over z+i\right) + {\alpha_2\over 2}
\ln\left(z-iq\over
z+iq\right).\eqn\uppmap$$
Lines of constant $\tau=X^+$ are equipotentials (where the electrostatic
potential is $\phi = \half(\rho + \bar \rho))$.   Close to the points $z=i$ and
$z=iq$ the equipotentials are circles in the $z$ plane, while near the real
axis the equipotentials are horizontal lines.  The electric field,
$\partial_\alpha \phi$,   is normal to any boundary (including the punctures)
and its integrated value around the $i$th boundary is the enclosed charge
$\alpha_i=  \oint_i \partial_n \phi d\sigma_t $  (where $_n$ denotes the
normal to the boundary).  The $\alpha_i$'s are arbitrary real constants subject
only to the condition $\sum_{i=1}^3\alpha_i=0$ (so that
$\alpha_3=-\alpha_1-\alpha_2$).   On a surface of genus zero
with $B$ boundaries there are generically $B-2$ zeroes of the electric field
(where a zero
on a boundary counts as half a zero).   In the present case  there is a
single zero  corresponding to a turning point, $z_0$,
of the map given by the solution of 
$${\partial \rho  \over \partial z} = 0 = { i \alpha_1\over z^2+1} +
{iq\alpha_2 \over z^2+q^2}. \eqn\turnzero$$
The solution is 
$$z_0^2 = -q\left({q+\eta \over 1 + \eta q} \right), \eqn\znought$$
where $\eta \equiv \alpha_2/\alpha_1$.  
The value of the modulus $q$ is determined by the value of $T$. 

 We shall choose $|\alpha_1| > |\alpha_2|$ (so that $|\eta| < 1$) with no loss
of generality but the properties of the map depend sensitively on the sign of
$\eta$.

(a)  {\it $\eta>0$}.  In this case $z_0^2$ is negative so that the zero lies on
the imaginary $z$ axis and $q\le |z_0| \le 1$.   The $\rho$ plane has the form
of two incoming (or outgoing) closed strings entering at $\tau=\infty$ (or
$\tau =-\infty$) and joining at the turning point $\rho_0 \equiv -T + i\sigma =
\rho(z_0)$  (which is real)  to form  a closed string of width
$-\pi(\alpha_1+\alpha_2)$ which evolves to the state $|X_3\rangle$ at $\rho
=0$ (\fone(a)).   The amplitude for this process is given by \procesr\  (with
legs 1 and 2 incoming) and as $q$ spans its complete range, $0\le q\le 1$, the
interaction point spans its complete range, $0\le T\le \infty$.

(b)  {\it $\eta<0$}.   This corresponds to the case in which one string (say,
the one of width $\pi \alpha_1$) is incoming and the other is outgoing. 
Now $z_0^2$ is either positive or negative depending on the value of $q$.  For
$q> -\eta$,  $z_0^2$ is again negative and $0\le |z_0| \le q$.  The string
diagram  is one describing an incoming string of width $\pi \alpha_1$ evolving
from $\tau = -\infty$ until it breaks at $\tau = -T< 0$ into a string of width
$\pi \alpha_2$ which evolves to $\tau=\infty$ and a string of width
$-\pi(\alpha_1 + \alpha_2)$ that evolves to the boundary at $\tau=0$
(\fone(bi)).  $T$ decreases as $q$ is reduced from $q=1$ until a critical value
$q_c = -\eta$ is reached where $z_0 =0$ so that the zero hits the real $z$ axis
and the interaction time is $T=0$.    In this range of $q$ the process is again
defined by \procesr\ (now with leg 2 outgoing), but this covers only part of
the complete moduli space.

 When $q$ is further reduced there are two real solutions to \znought\  so
there are two zeroes of the electric field  on the boundary -- the boundary has
two turning points.  These points $\rho^+ = \rho(z_0)= iR/2$ and $\rho^- =
\rho(-z_0) =-iR/2$ are on the imaginary $\rho$ axis ($\tau=0$) and the string
diagram is now one in which the boundary is represented by a vertical slit
(\fone(bii)) (so the parametrization of string 3 is double-valued).   This
situation was discussed in detail in  [\greena] in a theory with Neumann as
well as a Dirichlet boundaries.  The string diagram in this region of moduli
space is described by a new vertex that couples fields on legs 1 and 2 to a
field in the third Fock space that does not propagate in the usual manner.
This process may be expressed in the form
$$I'(1,2,X_3;R) = \langle W(R) |n_1\rangle
|n_2\rangle |X_3\rangle,\eqn\seconproc$$
where $\langle W(R)|$ is a new vertex describing this second term.  The
modulus, $R$,  changes from $a=\pi \alpha_1 + \pi
\alpha_2$ to $b= \pi \alpha_1$ as $q$ is reduced from $q_c$ to $0$.   In other
words, when $\eta <0$ the string diagram of \fone(bi) does not cover the whole
of  the moduli space for the world-sheet with boundary and the contribution of
\fone(bii) must be added.

Scattering amplitudes  are constructed by sewing world-sheets together and
integrating over $X_3$.  One
contribution to the four-string interaction is a $t$-channel \lq   pole term'
that is an integral over the two-dimensional moduli space,
$$\eqalign{&\int DX'_5 DX_5 \int_0^\infty dT \int_0^{2\pi} d\theta \  I^\dagger
(1,4,X'_5;0)\ \Delta
(X'_5, X_5;T) \ I(3,2, X_5;0)e^{-p^-T +i(N-\tilde  N) \theta} \cr & 
= \int_0^\infty  {dT\over \alpha}\int_0^{2\pi} d\theta \langle V_3|n_1\rangle
|n_4\rangle
|X_5\rangle \langle X_5|e^{(P^- -p^-)T+ i(N-\tilde N)\theta }|X_5'\rangle
\langle X_5'|\langle
n_3|\langle n_2| V_3\rangle, \cr}\eqn\onefour$$
where  the states 1 and 2 are incoming and states 3 and 4 are outgoing with
$\alpha_1 > -\alpha_4>0 $ and $-\alpha_3>\alpha_2>0$ (and $p^- = p_1^- +
p_2^-$).
In addition to the other pole terms,  the complete expression for the
four-particle amplitude requires the less familiar contact term  which arises
by sewing two $I'$ factors,
$$\eqalign{&\int DX_5 \int_a^b dR \ I^{\prime \dagger} (1,4,X_5;R)\
I^{\prime}(3,2,X_5;R)
\cr =&\int_a^b  dR\  \langle W(R) |n_1\rangle |n_4\rangle |X_5\rangle\ \langle
X_5|\langle n_2|\langle n_3|W (R)\rangle ,  \cr}\eqn\conti$$
where $a=\pi(\alpha_1+ \alpha_4)$,  $b = {\rm min} \{\pi\alpha_1,
-\pi\alpha_3\}$.
This contact interaction describes  two incoming closed strings (1 and 2)
touching at two points simultaneously (at $\sigma$ values separated by  $R$)
and interchanging string bits to emerge as the final strings (3 and 4).  It
contributes only on a one-dimensional sub-manifold of the two-dimensional
moduli space (the integrand  depends on only one real modulus,  $R$)  and is of
negligible weight in  the tree amplitudes of   bosonic closed-string theory.
Thus, mapping the four-string amplitude to  the complex $z$-plane, three of the
asymptotic  strings are associated with punctures that may be fixed at $z_1=0$,
$z_2=i$ and $z_3=\infty$ while the complex position of $z_4$ is the modulus
that is to be integrated.    The contact interaction arises from a portion of
the line $z_4=iy$ (where $y$ is real and $y<1$).  In superstring theories there
are operator insertions at the boundary turning points that can lead to
important contributions from the end-points  $R=a,b$ where the operators
collide \REF\klinka{J.  Greensite and F.R. Klinkhammer, {\it Superstring
amplitudes and
contact interactions} Nucl.  Phys. {\bf B304} (1988) 108.} [\klinka]  (see also
\REF\seiba{M.B.  Green and  N.
Seiberg,  {\it Contact interactions in superstring theory}, Nucl. Phys. {\bf
B299} (1988) 559.} [\seiba]).   Higher-order contact interactions arise in
amplitudes with more external particles on sub-manifolds of the
higher-dimensional moduli space appropriate to  such processes.  

Since \conti\  involves a sum over a complete set of states on leg 5 with no
intermediate propagator it  can be described as a contraction of an auxiliary
state, $|\Sigma\rangle$.  Such a state has no kinetic term
and its quadratic hamiltonian is just $\langle
I|\Sigma_1\rangle|\Sigma_2\rangle$, where $\langle I| $ is the identity.  The
vertex $\langle W(R)|$ gives a new interaction term,
$$H_3'   =  \int_a^b dR \langle W(R) |\Phi_1\rangle |\Phi_2\rangle
|\Sigma\rangle 
 = \int_a^b dR \ \langle W(R) |X_1\rangle|X_2\rangle |X_3\rangle\
\Phi[X_1] \Phi[X_2] \Sigma [X_3] , \eqn\auxinter$$
where $\Sigma[X] \equiv \langle X|\Sigma\rangle$ (and the limits $a$ and $b$
again depend on the values of the suppressed  integration variables $\alpha_1$
and $\alpha_2$).  The sum over intermediate
states in the second line  includes an integration over $\alpha_1$ and
$\alpha_2$ (with $\alpha_3$ determined by momentum conservation).

(c)  {\it The special case $\eta=-1$}.   In this case the incoming and outgoing
strings have $\alpha_1 = -\alpha_2$ so that the integral of the electric field
along the boundary vanishes.  Since the electric field, $\partial_n\phi$,  is
periodic this can only happen if it has an even number of zeroes (at least
two).  Indeed, since $z_0^2 =  q$,  there are precisely two zeroes on the real
$z$ axis for all values of $q$.  The process now   described  is a string of
width $\pi \alpha_1$ evolving from $\tau = -\infty$ to $\tau = \infty$ with a
vertical slit inserted at $\tau=0$ (\fone(c)).  In this case the boundary
carries no net $P^+$ component of momentum.  The turning points are given by
substituting $z_0^2 =q$ into \uppmap\ so that the length of the slit is given
by
$$ R = \pi \alpha_1 - 4 \alpha_1 \tan^{-1} (\sqrt q). \eqn\slitlen$$ 
 As $q$ varies from $q=1$ to $q=0$ this modulus varies from $R=0$ to $R = \pi
\alpha_1$.

A  parallel discussion can be given for open-string theories.   Again, the
usual three-string interaction does not cover the whole of moduli space and
there is a vertex coupling to an auxiliary open-string field,
$$\int_a^b dR  \langle w(R)|\psi_1\rangle |\psi_2\rangle
|\rho\rangle,\eqn\openfields$$
where $|\psi\rangle$ represents an open-string light-cone field and
$|\rho\rangle$ is the auxiliary open-string field.   The four-string contact
term that
arises by sewing two of these vertices together on the third leg is an integral
over $R$ as before but now this is of finite weight since the moduli space for
the tree amplitude with four external open strings is one-dimensional.   It is
the
familiar instantaneous interaction (occuring when two open strings touch at
internal points)  that arises in the light-cone description of the Veneziano
model
[\kakua].   The open-string diagram in which $\alpha_2=  - \alpha_1$
(analogous
to \fone(c)) is one in which the world-sheet is a strip of width $\pi
\alpha_1$.   The coordinates $X^i$ satisfy Neumann conditions on the
horizontal boundaries of the strip and the new vertex is represented by a
vertical section of the boundary of length $R\le \pi \alpha_1$ at $\tau=0$,
which has one turning point and on
which the coordinates take the values $X^i_3(\sigma)$.  

In addition, in any open-string field theory there is a vertex coupling a
closed string to two open strings.  In order to cover the whole of moduli space
in this case it is necessary to add a contribution from $\Sigma$, the
auxiliary closed-string field,  coupling to the interior of the world-sheet. 

\noindent{\it Vacuum expectation values}\hfill\break\indent
An interesting feature of the vertex shown in  \fone(c) (and the corresponding
open-string diagram) is that it makes a non-trivial contribution  even when the
momentum in leg 3 vanishes, in which case the total length of the  boundary is
$2R$.  This raises the possibility of modifying the usual theory by
assigning a non-zero expectation value to the zero-momentum auxiliary field so
that  \fone(c) would be a \lq tadpole' diagram that redefines the vacuum state.
   There are strong constraints on the form of this vacuum value due to the
requirement that the theory remain invariant under Poincar\'e transformations
generated by $J^{\mu\nu} = \int d\sigma P^{[\mu} X^{\nu]}$ and $P^\mu$.  

The idea is to couple $\Sigma$ to a source in order to shift the auxiliary
field in $H'_3$ to $\Sigma'$ defined by 
$$\Sigma  = \Sigma' + \lambda B \eqn\shiftfield$$
where $\lambda$ is an arbitrary constant weight and $B[X(\sigma)]$ is the
string field
expectation value.  
In the closed bosonic string theory there is an obvious candidate for this
expectation value that maintains Poincar\'e symmetry, namely, the
point-like field
$$B[X(\sigma)]\equiv \langle X|B\rangle = \delta^{D-2}[\partial_\sigma X^{i}
(\sigma)] ,
\qquad \eqn\boundaryy$$  
which has support only when $X^i(\sigma)$ is constant  (which further implies
that $X^-(\sigma)$ is constant).  
The state  $|B\rangle$  is defined by  $\partial_\sigma X^i(\sigma)|B\rangle
=0$ and $p^i|B\rangle=0$
(where $p^i$ is the  transverse momentum).  It may be expressed in terms of
modes as  [\greenaa] 
$$|B\rangle = \exp\left({\alpha^{i\dagger}_n \tilde \alpha^{i\dagger}_n\over
n}\right) |0\rangle \eqn\resboun$$ 
(where $\alpha_n$ and $\tilde \alpha_n$ are the usual modes of the closed
string and $|0\rangle$ is the zero-momentum ground state).   Upon shifting
$\Sigma$  in this way the auxiliary quantum field $ \Sigma'$
couples in $H_3'$ but in addition there is a new quadratic vertex,
$$\eqalign{H_2'  &=\lambda  \int_0^{\pi\alpha_1} dR \langle W(R)|\Phi_1\rangle
|\Phi_2\rangle |B\rangle\ \cr &
  = \lambda \int  DX_1^i DX_2^i DX^i_3 d\alpha_1 \int_0^{\pi \alpha_1} dR\cr
&\langle W(R)|X_1\rangle|X_2\rangle|X_3\rangle\
\Phi[X_1,\alpha_1]\Phi[X_2,-\alpha_1]\ \delta^{D-2} \left[\partial_\sigma
X_3^{i }\right],
\cr}\eqn\dirvertex$$
This describes the insertion of a zero-momentum boundary (of length $2R$) in
the world-sheet on
which the $X^\mu$ coordinates satisfy Dirichlet boundary conditions.  We need
to
check that $H_2 + H_2'$ is invariant under  Poincar\'e transformations of the
fields, at least to $O(\lambda)$.   The Lorentz transformations in the $D-2$
transverse directions, generated by $J^{ij}$, are obvious symmetries.
Transforming  the fields in   $H'_2$   by the non-linearly realized Lorentz
generator, $J^{i-}$,   results in a term proportional to  $J^{i-}|B\rangle =
P_0^-x^i|B\rangle$, where $P^-_0= \oint P^-(\sigma)$ is the integral of $P^-$
along a contour enclosing the boundary in \fone(c).   However, this contour may
be distorted into the sum of contours on legs 1 and 2.  The transformation of
$H'_2$ is then seen to cancel
with terms coming from a new $O(\lambda)$ variation of  fields in the
free-field term, $H_2$.  The closure of this Noether procedure has not been
checked at higher orders in $\lambda$. 

Furthermore, possible anomalies of the kind that
usually arise in the bosonic theory have been ignored.  In particular, the
singularities of $H'_2$ arising from the boundary of moduli space, $R=0$,  are
due to the closed-string dilaton and tachyon states coupling to the vacuum.
These are similar to the singularities of the usual theory with Neumann
boundaries, except that the relative sign of the dilaton and tachyon
singularities is reversed (in addition the trace of the graviton couples to the
Neumann boundary).  The singularity at $R=\pi \alpha_1$ is less familiar.  It
can be related to the presence of an open-string Lagrange multiplier state that
leads to a constraint on the  low-energy spectrum of the theory
\REF\greenc{M.B.  Green, {\it The influence of
world-sheet boundaries on critical closed string theory}, Phys.  Lett.  {\bf
B302}   (1993) 29.}[\greenc] that can only be properly interpreted after
summing over all iterations of the vertex.

\ifig\ftwo{(a) The closed-string propagator with iterated vacuum insertions
giving boundaries that are vertical slits at $\tau=\tau_i$.  (b)  The
open-string propagator. Vertical segments of the boundary  at $\omega=\omega_i$
represent interactions involving auxiliary open-string fields while closed
strings couple to the slits on the interior.  The boundary condition on all
boundaries is $\partial_\sigma X^{i}=0$.}
{\epsfbox{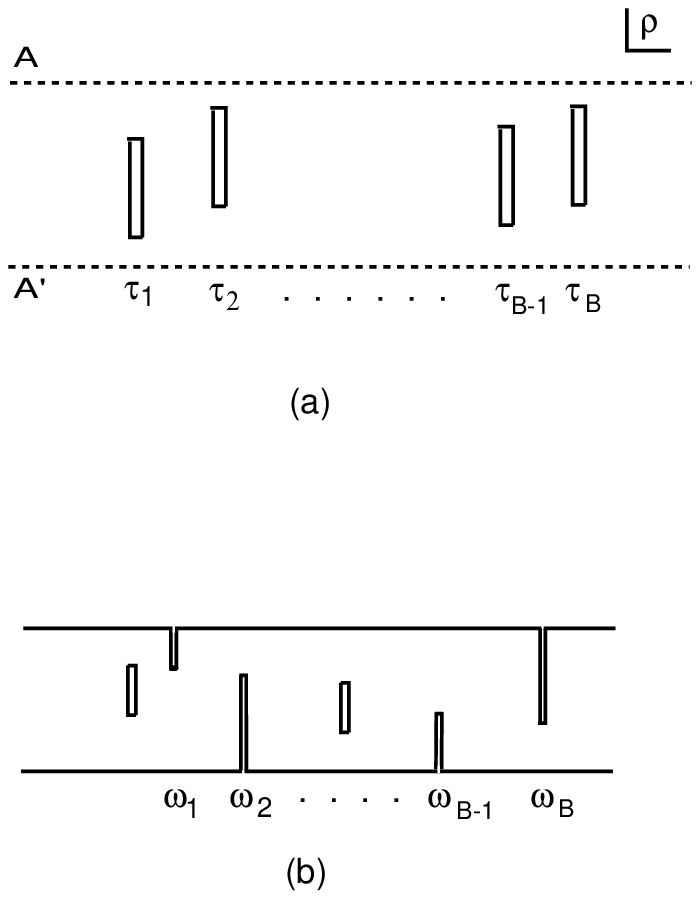}}
 The kinetic terms of the bare theory are modified by the presence of $H'_2$ so
that, in principle, the tree-level spectrum of the theory in this phase can be
determined by rediagonalizing the complete quadratic hamiltonian which is
equivalent to summing over all boundary insertions. Diagramatically, this
amounts to the iteration of the   vertex, leading to  \lq Dirichlet' string
theory in which the usual closed-string propagator is modified by a condensate
of such boundaries as in \ftwo(a).     Even though  summing over boundaries is
a formidable task it can be argued that  the resulting theory exhibits
point-like substructure induced by the presence of the boundaries which
introduce  a non-locality of the  mapping of the world-sheet into the target
space [\greena]  -- which was the original motivation for studying it.  

By contrast,   the insertion of a  Neumann boundary  is not described by a
local addition to the quadratic light-cone hamiltonian since it is not an
insertion at fixed $\tau=X^+$.   It might be viewed as a vacuum insertion in a
covariant treatment ( to be outlined below) in which $\tau$ is not identified
with the physical time coordinate.

The closed-string expectation value $\lambda |B\rangle$ also affects the
open-string theory since it modifies  the usual coupling between  a pair of
open strings and a closed string.  This gives a term quadratic in open-string
fields, illustrated by the vertical slits in \ftwo(b) in the interior of the
open-string strip.   The theory also possesses the usual Neumann boundary
insertions representing intermediate open-string states.
Since $\lambda$ is arbitrary there is the intriguing possibility of adjusting
its value so that the dilaton singularity cancels (at least to leading order in
string perturbation theory) in the sum of the Dirichlet and the Neumann
boundaries.

The situation with open-string theory also involves some new features.  The
obvious candidate for an open-string vacuum  value for the auxiliary field is
$\langle \rho [X]\rangle = b[X] =\langle X|b\rangle =
\delta^{D-2}[\partial_\sigma X^{i}(\sigma)]$,  so that
$$ \partial_\sigma X^i(\sigma) |b\rangle =0.\eqn\diropens$$
The open-string point-like state $|b\rangle$  (considered in \REF\greenv{M.B.
Green, {\it Locality and currents for the dual string},
Nucl. Phys.{\bf B103} (1976) 333.} [\greenv] in the context of the off-shell
states of non-zero momentum of  \REF\schwarza{J.H. Schwarz,  {\it Off-shell
dual amplitudes without ghosts},
{Nucl. Phys.} {\bf B65} (1973) 131.}\REF\fairliea{E.F. Corrigan and
D.B. Fairlie, {\it Off-shell states in dual resonance theory}, Nucl. Phys. {\bf
B91} (1975)
527.} [\schwarza, \fairliea])  is not a Lorentz scalar since the action of
$J^{i-}$ on the state is anomalous (it is a mixed angular momentum state of
zero helicity).  If this could be remedied,  the result of  the vacuum
modification to the open string would be the world-sheet in \ftwo(b), in which
 $\partial_\sigma X^{i}=0$ along the boundary so that the vertical segments
have
Dirichlet conditions and the horizontal Neumann.  The world-sheet may be mapped
to the upper-half plane with the boundary as the real axis which is composed of
segments on which the boundary conditions are alternately Dirichlet and
Neumann.   Summing over all possible insertions  results in an open string with
point-like energy densities at its end-points, whereas the expectation value
for the closed-string auxiliary field  generated point-like densities at
interior points.   Correspondingly, external (\lq flavour') currents can couple
to the point-like densities at the ends of the string whereas off-shell
closed-string currents can couple to interior point-like sections.

\noindent{\it  Covariant string field theory}\hfill\break\indent
The above treatment emphasized the light-cone gauge in which  perturbative
string field theory can be understood precisely.  However, the general idea
that open-string world-sheets  can be obtained by attaching tadpoles with
zero-momentum boundaries to closed world-sheets is a well-known feature of  the
conformal symmetry of string theory.   It should therefore be possible to
express the earlier results using a  covariant formulation of string field
theory such as that of 
\REF\wittena{E.  Witten, {\it Non-commutative geometry and string field
theory}, Nucl.  Phys.  {\bf B268} (1986) 253.} [\wittena] or the \lq
light-cone-like' formalism of  [\siegela--\neveua].   In the latter approach
many of the
earlier manipulations remain valid but  the expressions, that now include
Faddeev--Popov ghost
coordinates, are manifestly Poincar\'e-covariant -- it is  BRST invariance that
needs to be checked.   The arguments involving \fone\ show that contact terms
again arise for general values of $\alpha$.  However,  since $\alpha$ is no
longer a component of the string momentum the off-shell pole contributions in
\fone(a) and \fone(bi) can contribute even when the momentum in leg 3 is zero.
The end-state of the cylinder at $\tau = 0$ may therefore couple to the vacuum.
 The covariant expectation value of a string field, $B[X]$,  may now either
correspond to a
Dirichlet boundary, on which $X^{\mu \prime}(\sigma) |B\rangle_D =0$ (where
$\mu = 0,1,  \dots, D-1$)  or to a Neumann boundary,  on which
$P^\mu(\sigma)|B\rangle_N =0$.   These  boundary states are both BRST invariant
(after incorporating the appropriate ghost coordinates)
\REF\callana{C.G. Callan, C. Lovelace, C.R. Nappi and S.A. Yost,
{\it Adding holes and cross-caps to the superstring},
Nucl. Phys. {\bf B293} (1987) 83.}\REF\polchina{J.
Polchinski and Y. Cai, {\it Consistency of open superstring theories}, Nucl.
Phys. {\bf B296}
(1988) 91.}   [\callana,\polchina,\greenaa].   In other words, either  Neumann
or Dirichlet boundaries can arise as   vacuum values of the
closed-string field.  

The constraints of supersymmetry further limit the possibility of assigning
non-zero expectation values to the fields in superstring theories.   The
discussion in [\callana,\polchina] suggests that the usual type 1 theory
(in which the open-string sector has Neumann boundary conditions)  may be
expressed as a type 2b closed-string theory in
which the closed-string field acquires an expectation value  that preserves a
linear combination of world-sheet supersymmetries and is equal to  a very
particular sum of a Neumann boundary state and a cross-cap state.  However,  it
is not obvious that a Dirichlet version of  superstring theory exists since the
two types of point-like boundary states of the type 2b
theory transform  non-trivially (they are off-shell versions of the massless
scalar and pseudo-scalar states that form the end-states of a string
supermultiplet \REF\greensup{M.B.
Green {\it Point-like states for type 2b superstrings}, DAMTP/94-19;
hep-th/9403040.}  [\greensup]).

\refout
\bye